# CONTROL OF POLARIZATION HYSTERESIS TEMPERATURE BEHAVIOR BY INTERFACIAL SCREENING IN THIN FERROELECTRIC FILMS

Anna N. Morozovska[1,2], Eugene A. Eliseev[3], Ivan S. Vorotiahin[1], Maxim V. Silibin[4,5], Sergei V. Kalinin[6*] and Nicholas V. Morozovsky[1†],

[1] *Institute of Physics, National Academy of Sciences of Ukraine,*
*pr. Nauky 46, 03028 Kyiv, Ukraine*

[2]*Bogolyubov Institute for Theoretical Physics, National Academy of Sciences of Ukraine,*
*14-b Metrolohichna str. 03680 Kyiv, Ukraine*

[3]*Institute for Problems of Materials Science, National Academy of Sciences of Ukraine,*
*Krjijanovskogo 3, 03142 Kyiv, Ukraine*

[4] *The Center for Nanophase Materials Sciences, Oak Ridge National Laboratory,*
*Oak Ridge, TN 37831*

[5] *National Research University of Electronic Technology "MIET",*
*Moscow, Zelenograd, Russia*

[6]*Institute for Bionic Technologies and Engineering, I.M. Sechenov First Moscow State Medical University, 2-4 Bolshaya Pirogovskaya st., Moscow, Russia, 119991*

---

[*] corresponding author, e-mail: sergei2@ornl.gov
[†] corresponding author, e-mail: nicholas.v.morozovsky@gmail.com

**Abstract**

Ferroelectric interfaces are unique model objects for fundamental studies of polar surface properties such as versatile screening mechanisms of spontaneous polarization by free carriers and possible ion exchange with ambient media. The effect of ionic adsorption by electrically-open (i.e. non-electroded) ferroelectric surface on polarization reversal in the ferroelectric had been investigated experimentally and theoretically, however the effect influence on the temperature behaviour of polarization hysteresis remains unexplored. Also the comparative theoretical analysis of the ferroelectric hysteresis for linear and nonlinear electronic screenings, and strongly nonlinear ionic screening of the spontaneous polarization was absent.

In this work we study the free energy relief of a thin ferroelectric film covered by a screening charge layer of different nature and ultra-thin gap separating the film surface from the top electrode and calculate hysteresis loops of polarization and screening charge at different temperatures in the considered system. The dependence of the screening charge density on electric potential was considered for three basic models, namely for the linear Bardeen-type model of electronic surface states (BS), nonlinear Fermi-Dirac (FD) density of states describing two-dimensional electron gas at the film-gap interface, and strongly nonlinear Stephenson-Highland (SH) model describing the surface charge density of absorbed ions by Langmuir adsorption isotherms. Appeared that BS, FD and SH screening charges, which properties are principally different, determine the free energy relief (including polarization values corresponding to the free energy minima at different applied voltages) and control the shape of polarization hysteresis loops at different temperatures.

For BS screening polarization loops undergo classical second-order transition from the ferroelectric square-like shape to paraelectric curves with the temperature increase. The temperature dependence of the loops shape speaks in favor of the second-order ferroelectric phase transition for a thin $BaTiO_3$ film with BS screening, counter intuitively to the $BaTiO_3$ single-crystal that undergoes the first-order phase transition scenario. For FD screening we revealed the transition from a single ferroelectric-like to double antiferroelectric-like polarization hysteresis loops that happens with the temperature increase. Notably, antiferroelectric-like loops exist in a wide temperature range of about 100 K width and originate from the nonlinear electronic screening of ferroelectric polarization at the interface.

The transition from a single to double polarization hysteresis loop occurs for strongly nonlinear SH ionic screening at equal ion formation energies in Langmuir isotherms. For the case of different ion formation energies polarization reversal in the film is facilitated for one polarity and difficult for another polarity of applied voltage, which results into the shift and deformation of polarization hysteresis loops. The truncated and shifted minor loops of the ferro-ionic type open at room temperature in thin $BaTiO_3$ films. Minor loop gradually disappears with the temperature increase. With the film thickness increase polarization loops acquire slightly deformed and shifted ferroelectric-like shape at room and lower temperatures. The narrowing, horizontal shift, distortion and truncation of the loops occur with the temperature increase from 300 K to 500 K.

Comparing results for BS, FD and SH models of the interfacial screening, we conclude that the most versatile temperature behavior of polarization and screening charge hysteresis loops is inherent to SH model. It is conditioned by either symmetric or the asymmetric step-like dependence of the ionic charge density, described by Langmuir adsorption isotherms with different or equal ion formation energies, respectively. Obtained results open the way for control of polarization hysteresis at different temperatures by interfacial screening in thin ferroelectric films.

## I. Introduction

Electrically-open (i.e. non-electroded) ferroelectric interfaces are unique model objects for fundamental studies of polar surface properties such as versatile screening mechanisms of spontaneous polarization by free carriers and possible ion exchange with ambient media [1, 2].

As a matter of fact the conservation of spontaneous polarization in thin ferroelectric films usually results from either complete screening of polarization by ideally conducting electrodes, or from the emergence of the multidomain states in order to minimize the depolarization energy [3, 4, 5] in the case of incomplete screening by semiconducting electrodes, ultra-thin dead layers or physical gaps [6], leading to the non-zero spatial separation between polarization and screening charges [7, 8, 9, 10, 11]. The physical origin of natural dead layers and ultra-thin gaps is the unavoidable contamination of electrically-open ferroelectric surface that becomes paraelectric and provides maximally efficient electric screening of the polarization bound charge [12, 13].

Because of the long-range nature of depolarization effects the incomplete screening of ferroelectric polarization in the presence of ultra-thin dead layers and gaps leads to nontrivial domain structure dynamics [14]. These in turn causes unusual phenomena near the electrically opened surfaces and interfaces such as correlated polarization switching, formation of flux-closure domains in multiaxial ferroelectrics [15, 16, 17, 18], domain wall broadening in uniaxial [19, 20] and multiaxial ferroelectrics [17, 18]. Further examples of these behaviours include the crossover between different screening regimes in ferroelectric films [21, 22] and p-n junctions induced in 2D-semiconductors [23] separated by the ultra-thin gap from the moving ferroelectric domain walls. Notably the thickness of dead layer or gap should be small enough and its dielectric permittivity value should be high enough to prevent the dielectric breakdown when the voltage is applied to the film electrodes [24, 25, 26].

Most of the early papers, which consider the polar properties of ferroelectrics with electrically-open surfaces and interfaces, totally ignored nonlinear character of screening charges inherent to all self-screening mechanisms. Recent examples of linear screening consideration can be found in Refs. [17, 18, 27], where the authors studied the polar properties of thin ferroelectric films and nanoparticles covered by screening layer that's charge density $\sigma(\varphi)$ is linearly proportional to the electric potential $\varphi$ in the framework of Bardeen model [28] of the electronic surface states (**BS**), namely $\sigma(\varphi) \sim \varphi/L_S$. It was established that the critical thickness of the film and its transition temperature to a paraelectric phase strongly depend on the screening length $L_S$ as well as the joint action of the BS screening and flexoelectric coupling can remarkably modify polar and electromechanical properties of thin ferroelectric films. Effects similar to BS screening are expected for screening by electrodes with finite density of states [29].

Probably one of the most studied nonlinear models of the screening charge layer is the Fermi-Dirac (**FD**) density of states corresponding to the electron gas in a two-dimensional (2D) semiconductor (see e.g. Refs. [30, 31] for a single-layer graphene on a ferroelectric film). The FD charge density $\sigma(\varphi) \sim \sinh(e\varphi/k_B T)$. The mutual impact of the FD screening charge on the domain structure dynamics in a ferroelectric film is considered theoretically in Refs.[23, 32]. As a result on nonlinear screening nontrivial voltage and temperature behaviour of the ferroelectric polarization and 2D-carrier concentration dynamics was revealed in the nanostructure "graphene channel on $Pb_{0.55}Zr_{0.48}TiO_3$ film with domain walls". The transition from a single ferroelectric-like to double antiferroelectric-like polarization hysteresis loop has been predicted in the system with the temperature increase. Notably, antiferroelectric-like polarization loops exist in a wide temperature range (350 – 500) K.

A significantly more interesting example of the strongly nonlinear model describing interface charge is Stephenson and Highland (**SH**) model [1, 33]. Stephenson and Highland consider an ultrathin ferroelectric film in equilibrium with a chemical environment that supplies at least 2 different charge species "*i*" (e.g. ions and vacancies) to compensate its polarization bound charge at the interface. The φ-dependence of absorbed ions charge density is step-like, $\sigma_i(\varphi) \sim \left(1 + q_i \exp\left(\frac{\Delta G_i^{00} + eZ_i\varphi}{k_B T}\right)\right)^{-1}$ and can be either odd or asymmetric depending on the relation between the ions formation energies $\Delta G_i^{00}$ in Langmuir adsorption isotherms. Within SH model the screening by ions is electrically coupled to the electrochemical processes at the interface and thus the stabilization of ferroelectric state in ultrathin $PbTiO_3$ films occurs due to the chemical switching [34, 35, 36]. This coupling between ionic electric field and electric polarization in the film results into non-trivial effects on ferroelectric phase stability and phase diagrams [1, 33]. In general case the polar state of ferroelectric film in contact with atmosphere is undefined due to the presence of mobile electrochemically active and physically sorbable components at the interface [37, 38].

Recently, we modified the SH approach [33] allowing for the presence of the gap between the ferroelectric surface covered by ions and the SPM tip [39, 40, 41]. The analysis [39 – 41] leads to the elucidation of the **ferro-ionic, anti-ferroic** and **electret-like non-ferroelectric** states, which are the result of nonlinear electrostatic interaction between the single-domain ferroelectric polarization and absorbed ions with the charge density described by Langmuir adsorption isotherm. The properties of ferro-ionic states were described by the system of coupled equations. Further we study the stability of ferro-ionic states with respect to the domain structure formation and polarization reversal scenarios and reveal unusual dependences of the film polar state and domain structure properties on the ion formation energies, their difference, and applied voltage [42]. The electric field-induced phase transitions into ferro-ionic and anti-ferroionic states are possible in thin films covered with ion layer of electrochemically active nature.

We can conclude from the above review that the impact of ionic adsorption by electrically-open ferroelectric surface on polarization reversal in the ferroelectric had been investigated experimentally [1, 33 - 36, 39, 43, 44, 45, 46] and theoretically [1, 33, 39 - 42], however the effect influence on the temperature behaviour of polarization hysteresis remains unexplored. Also the comparative theoretical analysis of the ferroelectric hysteresis for linear and nonlinear electronic screenings, and strongly nonlinear ionic screening of the polarization was absent.

To fill the gap in the knowledge, in this work we study the free energy relief of a thin ferroelectric film covered by a screening charge layer of different nature and ultra-thin gap separating the film surface from the top biased electrode. We calculate hysteresis loops of polarization and screening charge at different temperatures. The dependence of the screening charge density on electric potential was considered for three basic models, namely for the linear BS electronic surface states, nonlinear FD density of states describing two-dimensional electron gas at the film-gap interface, and strongly nonlinear SH model describing the surface charge density of absorbed ions by Langmuir adsorption isotherms.

The manuscript is structured as following. Basic equations with boundary conditions and analytical expressions for the BS, FD and SH charge densities are discussed in **section II**. Free energy of the system is considered in **section III**. The impact of the surface screening nature (on example of the basic screening models) on thermodynamics of the polar and charge states is analyzed in **section IV.A.** The influence of the screening model on the quasi-static hysteresis loops of polarization and screening charge at different temperatures is discussed in **section IV.B. Section V** is a brief discussion and conclusions. Electrostatic problem, description of physical variables and their numerical values are given in **Appendixes A** and **B**, respectively.

## II BASIC EQUATIONS AND MODELS OF THE SCREENING CHARGES 2D-DENSITY

Let us we consider the system consisting of electron conducting bottom electrode, ferroelectric film covered by screening layer with 2D charge density $\sigma(\varphi)$, ultra-thin gap (or dead layer) separating the electrically-open surface of the film from the top electrode (either ion conductive planar electrode or flatted apex of SPM tip), as shown in **Fig. 1(a)**. The film-gap interface provides direct exchange of screening charges with ambient media. Mathematical statement of the problem is listed in Refs. [40, 41, 42], as well as in **Appendix A.**

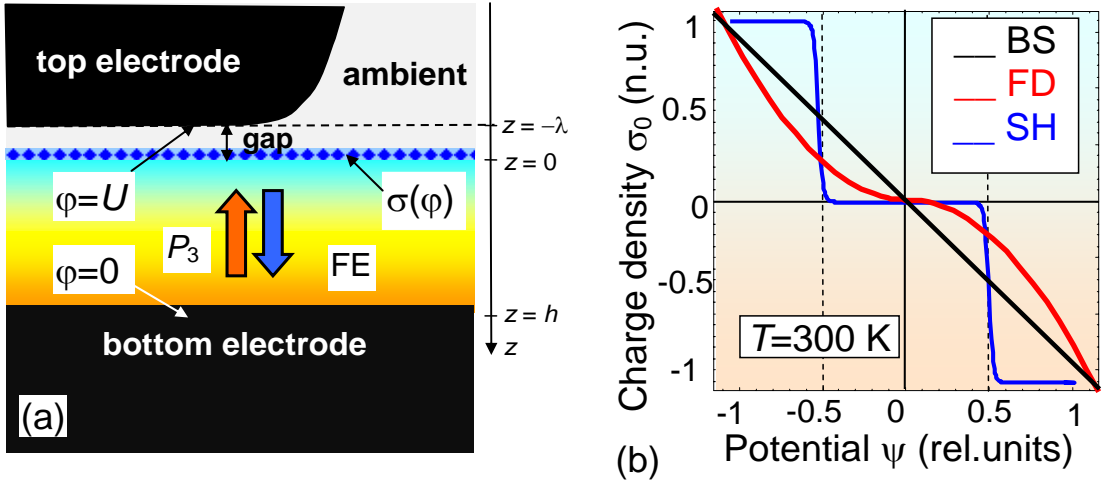

**FIGURE 1**. (a) Layout of the considered system, consisting of electron conducting bottom electrode, ferroelectric (FE) film, screening layer with 2D-charge density $\sigma(\varphi)$, ultra-thin gap separating the film surface and the top electrode. (b) Schematic dependence of the 2D charge densities $\sigma_0$ vs. acting potential $\psi$ calculated for Bardeen-type surface states (black line "BS"), Fermi-Dirac electron gas (red curve "FD"), and Stephenson-Highland (blue step-like curve "SH") screening model.

Since we account for the presence of the ultra-thin dielectric gap (or dead layer) between the top electrode and the ferroelectric surface, the linear equation of state $\mathbf{D} = \varepsilon_0 \varepsilon_d \mathbf{E}$ relates the electrical displacement $\mathbf{D}$ and electric field $\mathbf{E}$ in the gap. Here $\varepsilon_0$ is a universal dielectric constant and $\varepsilon_d \sim (1-100)$ is the relative permittivity of the physical gap or dead layer. A wide band-gap ferroelectric film can be considered dielectric. Quasi-static electric potential $\varphi$ satisfies electrostatic equations for each layer. Namely $\Delta\varphi = 0$ in the gap and $\left(\varepsilon_{33}^b \frac{\partial^2}{\partial z^2} + \varepsilon_{11}^f \Delta_\perp\right)\varphi = \frac{1}{\varepsilon_0}\frac{\partial P_3^f}{\partial z}$ in a ferroelectric film (see **Appendix A**). The boundary conditions (**BCs**) for the system are the equivalence of the electric potential to the voltage $U$ applied to the top electrode (or SPM tip apex) modeled by the flat region $z = -\lambda$; and the equivalence of the normal component of electric displacements to the screening charge density $\sigma[\varphi(\vec{r})]$ at the interface $z = 0$; the continuity of electric potential and normal component of displacements $D_3 = \varepsilon_0 E_3 + P_3$ and $D_3 = \varepsilon_0 \varepsilon_d E_3$ at gap - ferroelectric interface $z = 0$; and zero potential at the bottom conducting electrode $z = h$ [see **Fig. 1(a)**].

The polarization components of uniaxial ferroelectric film depend on the inner electric field $E$ as $P_1 = \varepsilon_0(\varepsilon_{11}^f - 1)E_1$, $P_2 = \varepsilon_0(\varepsilon_{11}^f - 1)E_2$ and $p_3 = P_3 + (\varepsilon_{33}^b - 1)E_3$, where background permittivity $\varepsilon_{33}^b \leq 10$ [12]. The dielectric permittivity $\varepsilon_{33}^f$ is related with the ferroelectric polarization $P_3$ via the soft mode. The evolution and spatial distribution of the ferroelectric polarization $P_3$ is given by the time-dependent LGD equation [40-42]:

$$\Gamma \frac{\partial P_3}{\partial t} + \alpha P_3 + \beta P_3^3 + \gamma P_3^5 - g \frac{\partial^2 P_3}{\partial z^2} = E_3, \tag{1}$$

In Eq.(1), the $\Gamma$ is kinetic coefficient; $\alpha = \alpha_T(T_C - T)$, $\beta$ and $\gamma \geq 0$ are the coefficients of LGD potential $G(P_i, U)$ expansion on the higher polarization powers [47]; $T$ is the absolute temperature, $T_C$ is Curie temperature. The **BCs** for polarization at the film surfaces $z = 0$ and $z = h$ are of the third kind $\left(P_3 \mp \Lambda_\mp \frac{\partial P_3}{\partial z}\right)\bigg|_{z=0,h} = 0$ and include extrapolation lengths $\Lambda_\pm$ [48, 49] at the film surfaces.

To describe the screening charge dynamics, we propose a relaxation equation [40, 41], $\tau \frac{\partial \sigma}{\partial t} + \sigma = \sigma_0[\varphi]$, where $\tau$ is the relaxation time. Expression for **the equilibrium 2D charge density** $\sigma_0[\varphi(\vec{r})]$ is considered for three basic models

**I. Bardeen-type density of surface states (BS).** Here, using Bardeen model [28] with corresponding approximations, we consider the special case of the screening charges which density linearly depends on electric potential $\varphi$, namely

$$\sigma_0(\varphi) = -\varepsilon_0 \varphi / L_S, \tag{2a}$$

where $L_S$ is the effective screening length introduced in Refs. [17, 18, 27]. The schematic dependence $\sigma_0(\varphi)$ is shown in **Fig. 1(b)** by black line.

**II. Fermi-Dirac (FD) density of states** corresponding to the 2D electron gas. The electronic gas is characterized by the charge density, $\sigma_0(\varphi) = e(p_{2D}(\varphi) - n_{2D}(\varphi))$, that is the difference of 2D concentration of electrons in the conduction band $[n_{2D}(\varphi) = \int_0^\infty d\varepsilon\, g_n(\varepsilon) f(\varepsilon - E_F - e\varphi)]$ and holes in the valence band $[p_{2D}(\varphi) = \int_0^\infty d\varepsilon\, g_p(\varepsilon) f(\varepsilon + E_F + e\varphi)]$, respectively. Electrons and holes density of states (DOS) are $g_n(\varepsilon)$ and $g_p(\varepsilon)$, $E_F$ is the position of Fermi energy level. The expressions for the gapless FD DOS are $g_n(\varepsilon) = g_p(\varepsilon) = 2\varepsilon/(\pi \hbar^2 v_F^2)$ (see e.g. Refs. [30, 31]). Using the expressions for $g_n(\varepsilon)$ and $g_p(\varepsilon)$ (characteristic for e.g. graphene charge density) and their Pade-exponential approximation derived in Ref.[23], the screening charge density acquires the form:

$$\sigma_0(\psi) = \frac{4(k_B T)^2 e}{\pi \hbar^2 v_F^2} \sum_{m=1}^{\infty} \frac{(-1)^m}{m^2} \sinh(m\psi) \approx \frac{2(k_B T)^2 e}{\pi \hbar^2 v_F^2} \left( \frac{1}{\eta(\psi)} - \frac{1}{\eta(-\psi)} \right), \quad (2b)$$

where $v_F$ is the Fermi velocity of the carrier, the functions are $\psi = \frac{e\varphi + E_F}{k_B T}$ and $\eta(\psi) = \exp(\psi) + 2\left(\psi^2 + \frac{\psi}{2} + \frac{2\pi^2}{12 - \pi^2}\right)^{-1}$. The schematic dependence of the density (2b) on the dimensionless voltage $\psi$ is shown in **Fig. 1(b)** by red curve.

**III. Stephenson-Highland (SH) model [33],** that is analogous to the Langmuir adsorption isotherm used in interfacial electrochemistry for adsorption onto a conducting electrode exposed to ions in a solution [50]. The dependence of equilibrium charge density $\sigma_0[\varphi]$ on electric potential $\varphi$ is controlled by the concentration of surface ions $\theta_i(\varphi)$ at the interface z = 0 in a self-consistent manner via Langmuir adsorption isotherms:

$$\sigma_0[\varphi] = \sum_i \frac{eZ_i \theta_i(\varphi)}{A_i} \equiv \sum_i \frac{eZ_i}{A_i} \left(1 + q_i \exp\left(\frac{\Delta G_i^{00} + eZ_i \varphi}{k_B T}\right)\right)^{-1}, \quad (2c)$$

where $e$ is an elementary charge, $Z_i$ is the charge of the surface ions/electrons, $1/A_i$ is the saturation densities of the surface ions, at that $i \geq 2$ to reach the charge compensation. The factor $q_i = \left(\frac{p_{atm}}{p_{exc}}\right)^{1/n_i}$, where $p_{exc}$ is the partial pressure of ambient gas relative to atmospheric pressure $p_{atm}$, $n_i$ is the number of surface ions created per gas molecule, $\Delta G_i^{00}$ is the standard free energy of the surface ion formation at $p_{exc} = 1$ bar and $U = 0$. Below we consider the case $p_{exc} = p_{atm}$.

Schematic step-like dependence of the screening charge density $\sigma_0$ on the dimensionless potential $\psi = \frac{e\varphi}{k_B T}$ calculated at $\Delta G_1^{00} = \Delta G_2^{00}$ is shown in **Fig. 1(b)** by blue curve. The screening charge is small or zero at $\psi = 0$, and then is abruptly increases. Note that the abrupt step-like dependence is left-shifted with respect to $\psi = 0$ for $\Delta G_1^{00} < \Delta G_2^{00}$, symmetric for $\Delta G_1^{00} = \Delta G_2^{00}$ and right-shifted for $\Delta G_1^{00} > \Delta G_2^{00}$ (see Refs.[40, 41, 42]).

Noteworthy, the developed solutions of Eq.(1) are sensitive to the thermodynamic parameters of corresponding screening charges given by Eqs.(2) [32, 51].

### III. FREE ENERGY OF THE SYSTEM FOR DIFFERENT SCREENINGS

In thermodynamic equilibrium, the Landau-Ginzburg-Devonshire (LGD) free energy is the sum of bulk ($G_V$) and surface ($G_S$) parts of the Gibbs free energy of a ferroelectric film, and the energy of the electric field outside the film ($G_{ext}$):

$$G = G_V + G_S + G_{ext}, \quad (3a)$$

$$G_V = \int_{V_{FE}} d^3r \left( \frac{\alpha}{2} P_3^2 + \frac{\beta}{4} P_3^4 + ... + \frac{g_{33ij}}{2} \left( \frac{\partial P_3}{\partial x_i} \frac{\partial P_3}{\partial x_j} \right) - P_3 E_3 - \frac{\varepsilon_0 \varepsilon_b}{2} E_i E_i \right), \quad (3b)$$

$$G_S = \int_S d^2r \left( \frac{\alpha_S}{2} P_3^2 - \frac{\varphi}{2} \sigma_0[\varphi] \right), \qquad G_{ext} = -\int_{\vec{r} \notin V_{FE}} d^3r \frac{\varepsilon_0 \varepsilon_e}{2} E_i E_i. \quad (3c)$$

Here $P_i$ are the polarization vector components. The tensor $a_{ij}$ is positively defined for linear dielectrics, and explicitly depends on temperature $T$ for ferroelectrics and paraelectrics. Below we use an isotropic approximation for the tensor coefficients $\alpha = \alpha_T (T - T_C)$, where $T$ is the absolute temperature, $T_C$ is the Curie temperature. All other tensors included in the free energy (3) are supposed to be temperature independent. Tensor $g_{ijkl}$ determines the magnitude of the gradient energy, and is also regarded positively defined. $\varepsilon_0$ is the vacuum permittivity, $\varepsilon_b$ is a relative background dielectric permittivity [12]. Extrapolation lengths $\Lambda_\pm = g_{33}/\alpha_S^\pm$

Here, we develop simplified analytical model to get insight into numerically analyzed behaviors of different polarization states. Since the stabilization of single-domain polarization in ultrathin perovskite films takes place due to the chemical switching (see e.g. Refs.[1, 33, 34, 35, 36]), we can assume that polarization distribution $P_3(x, y, z)$ is smooth enough. In this case, the behavior of the polarization averaged over film thickness $P = \langle P_3 \rangle$ and screening charge density $\sigma_0[\varphi]$ can be described via the coupled nonlinear algebraic equations derived in Refs. [40, 41]. For a given screening charge model the equations for polarization and overpotential are:

$$\alpha_R P + \beta P^3 + \gamma P^5 = \frac{\Psi}{h}, \quad (4a)$$

$$\Psi = \frac{\lambda(\sigma_0[\Psi] - P) + \varepsilon_0 \varepsilon_d U}{\varepsilon_0 (\varepsilon_d h + \lambda \varepsilon_{33}^b)} h. \quad (4b)$$

The physically justified free energy $G$, which formal minimization, $\frac{\partial G[P, \Psi]}{\partial P} = 0$ and $\frac{\partial G[P, \Psi]}{\partial \Psi} = 0$, gives the coupled equations (4) for polarization dynamics, has the form [42]:

$$\frac{G[P, \Psi]}{S} = h\left( \frac{\alpha_R}{2} P^2 + \frac{\beta}{4} P^4 + \frac{\gamma}{6} P^6 \right) - \Psi P - \varepsilon_0 \varepsilon_{33}^b \frac{\Psi^2}{2h} - \frac{\varepsilon_0 \varepsilon_d}{2} \frac{(\Psi - U)^2}{\lambda} + \int_0^\Psi \sigma_0[\phi] d\phi \quad (5)$$

Coefficient $\alpha_R = \alpha_T(T_C - T) + \dfrac{g_{33}}{h}\left(\dfrac{1}{\Lambda_+} + \dfrac{1}{\Lambda_-}\right)$ is the coefficient $\alpha$ renormalized by "intrinsic" gradient-correlation size effects (the term $\sim g_{33}/h$). The first term in Eq.(5) is LGD polarization energy. The second term, $\Psi P$, represents the energy of interaction of polarization $P$ with overpotential $\Psi$. The terms $\varepsilon_0\varepsilon_{33}^b\dfrac{\Psi^2}{2h}$ and $\dfrac{\varepsilon_0\varepsilon_d}{2}\dfrac{(\Psi-U)^2}{\lambda}$ are the energies of electric field in the ferroelectric film [where $\varphi = (h-z)\dfrac{\Psi}{h}$] and in the gap [where $\varphi = U - \dfrac{z+\lambda}{\lambda}(U-\Psi)$], correspondingly. The last term, $\int_0^\Psi \sigma_0[\phi]d\phi$, is the screening charge energy that's integration over $\Psi$ for the considered models yields

$$\Sigma[\Psi] \equiv \int_0^\Psi \sigma_0[\phi]d\phi = \begin{cases} -\varepsilon_0\dfrac{\Psi^2}{2L_S}, & BS \\ \dfrac{4(k_BT)^2 e}{\pi\hbar^2 v_F^2}\sum_{m=1}^\infty \dfrac{(-1)^m}{m^3}(\cosh(m\Psi)-1), & FD \\ \sum_i \dfrac{Z_i}{A_i}\left(e\Psi + k_BT\ln\left[\dfrac{1+q_i\exp(\Delta G_i^{00}/k_BT)}{1+q_i\exp((\Delta G_i^{00}+eZ_i\Psi)/k_BT)}\right]\right), & SH \end{cases} \quad (6)$$

The energy given by Eq. (5) has an absolute minimum at high $\Psi$. According to the Biot's variational principle [52], we can further use the incomplete thermodynamic potential that's partial minimization of over $P$ will give the coupled equations of state, and, at the same time, it has an absolute minimum at finite $P$ values. For the considered BS, FD and SH models we will analyze graphically the relief of the energy (5)-(6) along with its extremals given by Eqs (4).

For linear BS model Eq.(2a) the substitution of the extremal (4b) in expressions (5) and (6) yields

$$G[P,U] = \begin{pmatrix} \dfrac{\varepsilon_d U}{\varepsilon_d h + \lambda\varepsilon_{33}^b}P + \left(\dfrac{\lambda}{\varepsilon_0(\varepsilon_d h + \lambda\varepsilon_{33}^b)} + \alpha_R\left[1 + \dfrac{\lambda}{(\varepsilon_d h + \lambda\varepsilon_{33}^b)}\dfrac{h}{L_S}\right]\right)\dfrac{P^2}{2} \\ + \beta\left(1 + \dfrac{\lambda}{(\varepsilon_d h + \lambda\varepsilon_{33}^b)}\dfrac{h}{L_S}\right)\dfrac{P^4}{4} + \gamma\left(1 + \dfrac{\lambda}{(\varepsilon_d h + \lambda\varepsilon_{33}^b)}\dfrac{h}{L_S}\right)\dfrac{P^6}{6} \end{pmatrix} \quad (7)$$

As one can see the coefficients in Eq.(7) are renormalized by the terms proportional to $\dfrac{h}{L_S}$. Simple analytical expressions like (7) are hardly possible for FD and SH models.

## IV. POLAR AND CHARGE STATES FOR BASIC SCREENING MODELS

Below we will analyze graphically the free energy relief along with its extremals given by Eqs.(4) for BS, FD and SH models of the screening charges. The consideration of different screening charge mechanisms leads to the notable differences in the dependences of the overpotential and free energy on polarization. Consequently the hysteresis loops of polarization reversal and screening charge density strongly depend on the model, as explained below.

### A. Impact of the surface screening nature on the system thermodynamics

Since BS, FD and SH models are distinguished by screening charge carriers nature, the expressions (2) for their density $\sigma_0(\varphi)$ dependences on electric potential are distinguished by nonlinearity degree. Obtained results, presented in **Figs.2-4**, show that BS (**Fig. 2**), FD (**Fig. 3**) and SH (**Fig. 4**) models, and thereby chosen screening mechanisms, determine the shape of the free energy relief, namely the free energy cross-sections at fixed voltage and voltage-dependent free energy minima at certain polarization values. Difference in the densities $\sigma_0(\varphi)$ result in the noticeable difference of the free energy relief maps at zero voltage $U=0$ [compare **Figs. 2(a), 3(a)** and **4(a)**], quantity and sharpness of free energy minima and character of their transformation under applied voltage [compare different curves in **Figs. 2(b), 3(b)** and **4(b)**].

For BS model one of the symmetric minima of $G(P,U)$ with almost flat potential well at $U = 0$ transforms to asymmetric and smooth one shifted away from zero-polarization position $P=0$ under $U$ increase [compare black, red, magenta and blue curves in **Fig. 2(b)**]. For FD-model one pronounced relatively sharp symmetric minimum of $G(P,U)$ with two inflection regions on it wings exists at $U = 0$ [see black curve in **Fig. 3(b)**]. Under $U$ increase the dependence of $G(P,U)$ on polarization $P$ at first transforms to asymmetric curve with two minima of different shape and one inflection region, and then, into the asymmetric curve with the side global minimum and two inflection regions located on the different sides from the vertical line $P=0$ [compare red, magenta and blue curves in **Fig. 3(b)**].

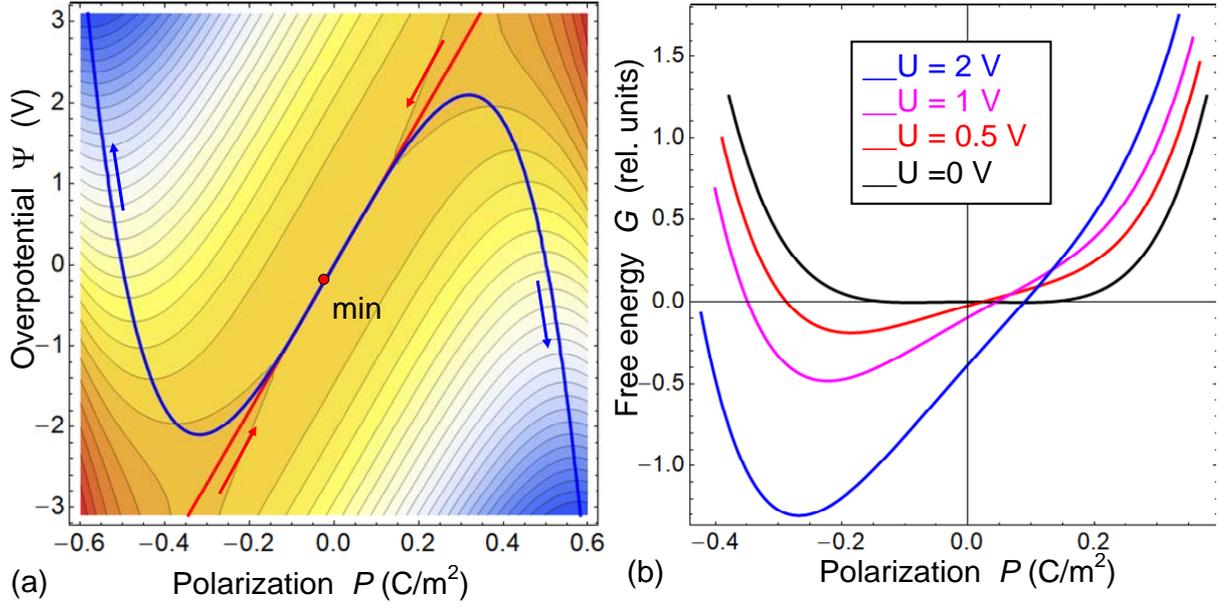

**FIGURE 2. Free energy relief for Bardeen model of screening charges.** (a) Free energy relief at room temperature for $BaTiO_3$ film of thickness 92 nm and zero applied voltage $U = 0$. Blue and red curves are extremals divergent and convergent towards absolute minimum, given by Eqs. (4a) and (4b), respectively. (b) Free energy profile at $U = 0$, 0.5, 1 and 2 V (black, red, magenta and blue curves). Screening length $L_S = 0.1$ nm.

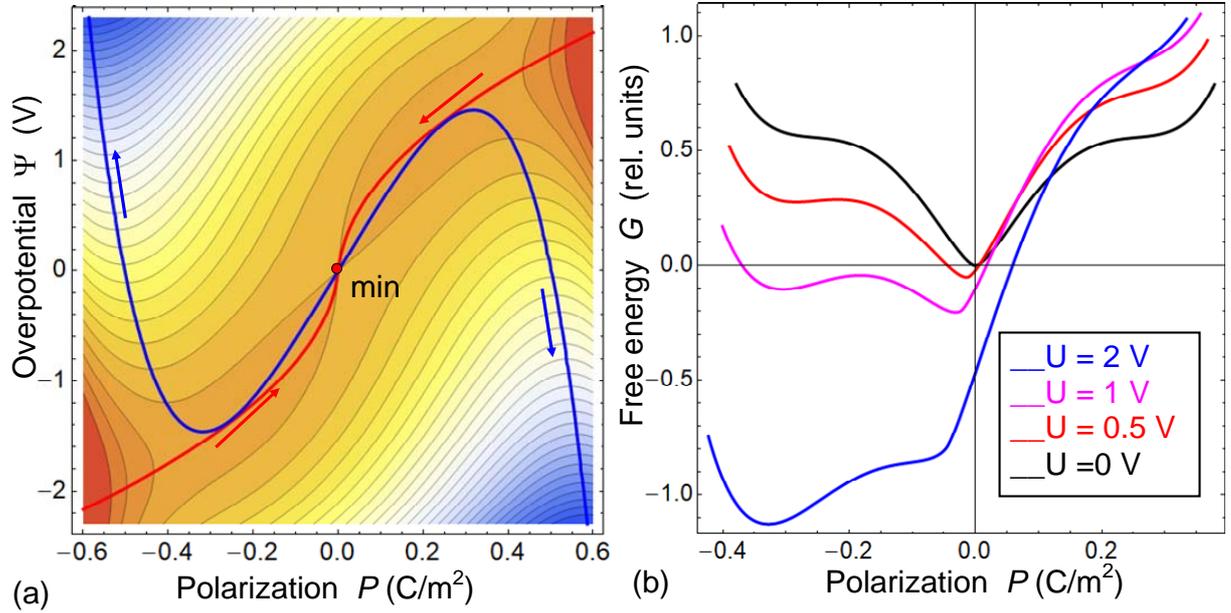

**FIGURE 3. Free energy relief for Fermi-Dirac model of screening charges.** (a) Free energy relief at room temperature for $BaTiO_3$ film of thickness 64 nm and zero applied voltage $U = 0$. Blue and red curves are extremals divergent and convergent towards absolute minimum, given by Eqs. (4a) and (4b), respectively. (b) Free energy profile at $U = 0$, 0.5, 1 and 2 V (black, red, magenta and blue curves). Fermi velocity $v_F = 10^6$ m/s.

For SH screening by ions with equal and relatively small formation energies ($\Delta G_1^{00} = \Delta G_2^{00} = 0.1$ eV), a symmetric "horn"-like dependence of $G(P, U=0)$ on polarization $P$ has a smoothed shallow minimum at $P=0$ [see **Fig. 4(d)**]. Under the voltage increase up to 0.5 V, the symmetric polarization dependence of $G(P,U)$ transforms into asymmetric one with two shallow minima (the side one and the former central one) and one inflection region, which are located on different sides from the line $P=0$. Under further voltage increase up to 2V, the small shift of the former central minimum away from $P=0$ takes place and pronouncedly nonlinear falling of $G(P,U)$ to negative values appears.

In the case of higher and equal ion formation energies ($\Delta G_1^{00} = \Delta G_2^{00} = 0.2$ eV) [see **Fig. 4(e)**], the voltage transformation of symmetric dependence of $G(P,U)$ at $U=0$ is similar to the one shown in **Fig. 4(d)**. At that, however, the side inflection region transforms in a shallow minimum and the former central shallow minimum transforms in the break with the voltage increase. The nonlinearity degree of voltage induced falling of the $G(P,U)$ dependence on polarization $P$ is a bit higher than in the previous case.

In the case of different and high ion formation energies ($\Delta G_1^{00} = 0.2$ eV and $\Delta G_2^{00} = 0.4$ eV) the dependence of $G(P,U)$ on polarization $P$ is asymmetric at $U=0$ and has sharper minimum in the vicinity of $P=0$, in contrast to the previous case [see **Fig. 4(e)** and compare with **Fig. 4(d)**]. Under voltage increase, this minimum shifts away from the former position at $P=0$. Similarly to the previous case, the side inflection region transforms in a shallow minimum and former central minimum transforms in the break. The nonlinearity degree of the voltage induced falling of $G(P,U)$ curve is almost the same as in the previous case.

Comparing **Figs 2-4** we should underline that the general tendency of $G(P,U)$ falling downwith the voltage increase is minimal for BS model (less than 0.5 rel. units at $P=0$ at $U=2V$), a bit higher for FD model (about 0.5 rel. units at $P=0$ at $U=2V$) and maximal for SH model (about 5 rel. units for at $P=0$ at $U=2V$).

To resume the subsection, various nature of the screening charges (electrons, holes or adsorbed ions) and thus the distinction of their localization and interaction with the film surface for considered BS, FD and SH models leads to different nonlinearity degree of their density $\sigma_0$ dependence on overpotential $\psi$ [see the dependences $\sigma_0(\psi)$ in **Fig. 1(b)** described by Eqs.(2)]. The coupling between polarization $P(U)$ and screening charge $\sigma_0(U)$ through the overpotential $\psi$ described by Eqs (4a, b) results in the hysteretic behaviour of the both quantities, $P(U)$ and $\sigma_0(U)$, and therefore leads to the different peculiarities of quasi-static $P(U)$ and $\sigma_0(U)$ hysteretic loops represented and analyzed in next subsection.

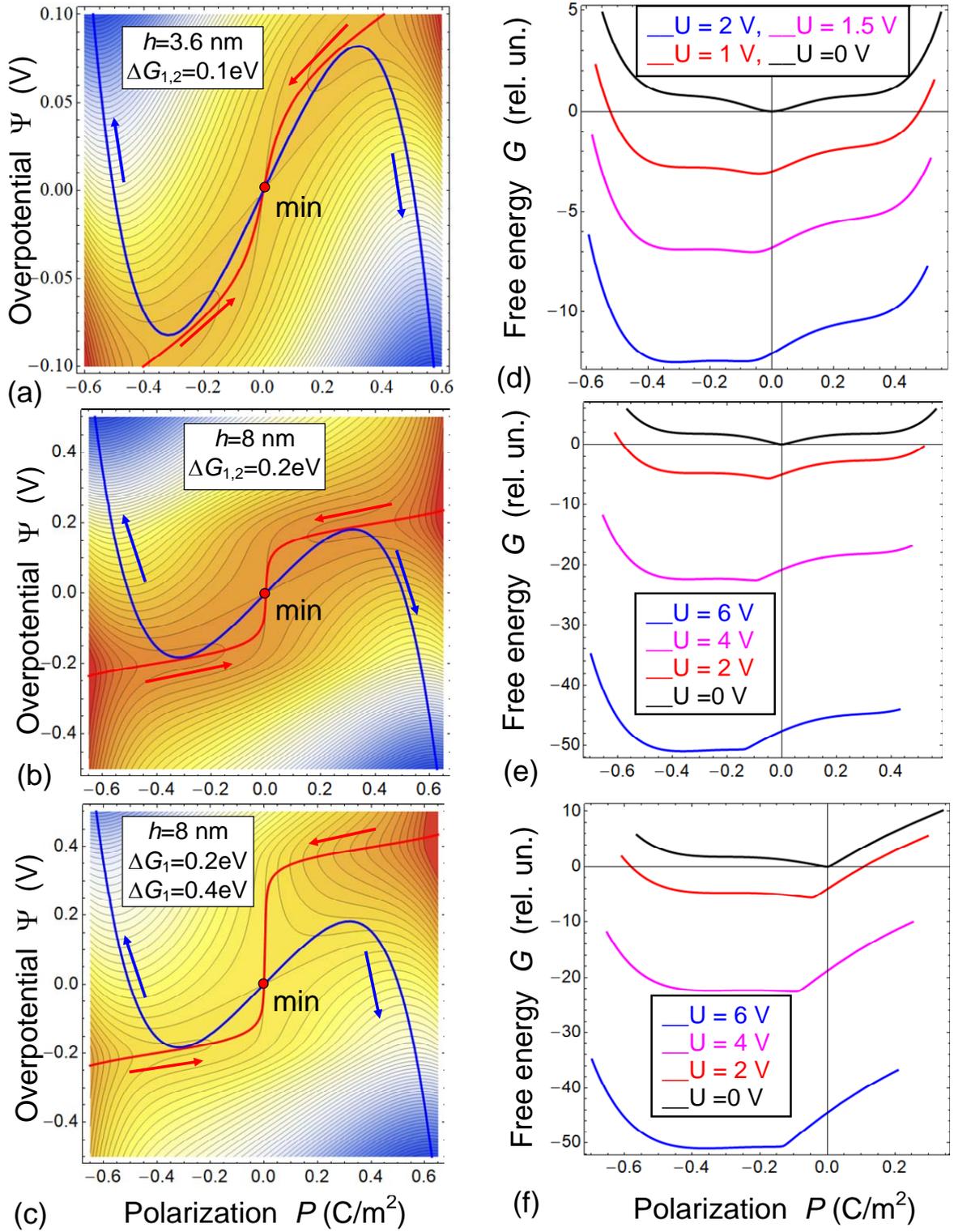

**FIGURE 4. Free energy relief for Stephenson-Highland model of ionic screening. (a, b, c)** Free energy relief at room temperature, zero applied voltage $U = 0$, BaTiO$_3$ film thickness 3.6 nm **(a)**, and 8 nm **(b, c)**. Blue and red curves are extremals divergent and convergent towards absolute minimum, given by Eqs.(4a) and (4b), respectively. **(d, e, f)** Corresponding profiles of the free energy at different voltages U = 0, 2, 4 and 6 V (black, red, magenta and blue curves) indicated at legends. Ion formation energies are $\Delta G_1^{00} = \Delta G_2^{00} = 0.1$ eV **(a, c)**; $\Delta G_1^{00} = \Delta G_2^{00} = 0.2$ eV **(b, d)**; $\Delta G_1^{00} = 0.2$ eV and $\Delta G_2^{00} = 0.4$ eV. **(c, f)**. Other parameters of SH model are listed in **Table SI** in **Appendix B**.

## B. Impact of the surface screening nature on the quasi-static hysteresis loops of polarization and charge

Firstly let us analyze the impact of the screening charges described by Bardeen surface states model on the quasi-static hysteresis of polarization and charge in a ferroelectric nanostructure shown in **Fig. 1(a).** Linear dependence of the BS charge density $\sigma_0$ on acting potential $\varphi$ is shown by black curve in **Fig. 1(b)**.

From **Figure 5**, for temperatures less than Curie temperature $T_C$, there is a hysteresis of ferroelectric type for $P(U)$ and the corresponding hysteresis $\sigma_0(U)$ with quasi-linear parts outside the hysteresis in a 100-nm BaTiO$_3$ film. The temperature increase leads to a simultaneous narrowing of the hysteresis loops $P(U)$ and $\sigma_0(U)$ up to their disappearance, since $P(U)$ and $\sigma_0(U)$ are strongly coupled via Eqs.(4). Loops disappearance is accompanied by the decrease of remanent polarization $P(0)$ from 0.27 C/m$^2$ to 0, and by the expansion of $\sigma_0(U)$ maxima and their escape from the hysteresis region beyond its limits, as well as by its shift to higher voltages, while the maximal values $\sigma_0(U)$ remains the same. The temperature behavior of $P(0)$ is similar to the behavior of the spontaneous polarization in the case of a second-order ferroelectric phase transition [53, 54]. However, for a chosen $T_C$ = 381 K of bulk BaTiO$_3$, $P(0) \neq 0$ even for 450 K. Therefore, the electric field of the screening charge unexpectedly strongly supports ferroelectric-like remanent polarization $P(0)$ above $T_C$. This corresponds to the known behavior of $P(T)$ under the second-order ferroelectric phase transition (while BaTiO$_3$ undergoes the first-order phase transition in the bulk), when an external electric field induces the polarization in a shallow paraelectric phase at temperatures somewhat higher than $T_C$ [30]. Allowing for the relationship between $P(U)$ and $\sigma_0(U)$, described by Eqs. (4a) and (4b), the behavior of the screening charge density $\sigma_0(U)$ for $T<T_C$ is determined by the hysteresis ferroelectric-like dependence $P(U)$, while the density $\sigma_0(U)$ is governed by the nonlinear paraelectric-like dependence $P(U)$ for $T>T_C$. Therefore the nonlinear dependence of $\sigma_0(U)$ in the vicinity of $T_C$ does not follow from the linear dependence of $\sigma_0(\varphi)$ shown in **Fig.1(b)** for the BS model. However the dependence $\sigma_0(U)$ is also linear in the region of a quasi-linear variation of polarization $P(U)$ in the range $-1V <U <1V$ for elevated temperature $T$ = 500 K. It should be noted that the temperature changes of the $\sigma_0(U)$ loop width are smooth and correlate with the changes of the loop $P(U)$ width. Finally, we did not calculate any polarization loops with a constriction or double hysteresis loops, speaking in favor of the second-order ferroelectric phase transition for BS screening of spontaneous polarization in thin BaTiO$_3$ film, counter intuitively to the thick BaTiO$_3$ single-crystal that undergoes the first-order phase transition scenario. The physical origin of the change in the order of phase transition is the joint action of BS screening and finite size effect.

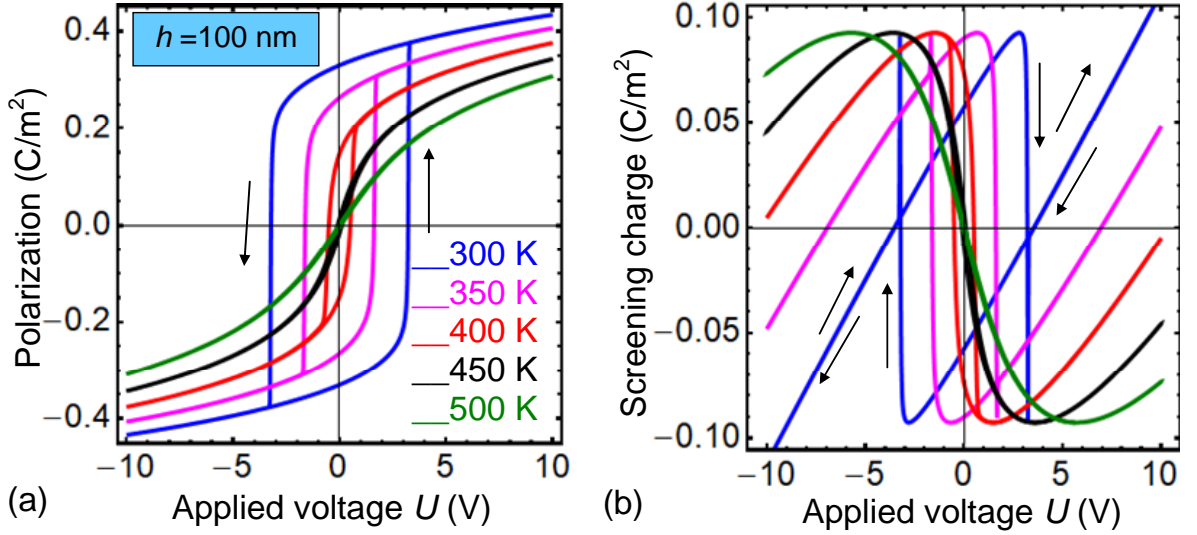

**FIGURE 5. Quasi-static hysteresis loops for Bardeen surface states model of the screening charge.** Quasi-static hysteresis loops of the average ferroelectric polarization $P(U)$ **(a)** and screening charge density $\sigma_0(U)$ **(b)** calculated for different temperatures 300, 350, 400, 450 and 500 K (different curves) and BaTiO$_3$ film thicknesses $h$ = 100 nm. Screening length $L_S = 0.1$ nm; $T_C$=381 K for a bulk BaTiO$_3$. Other parameters are listed in **Table SI** in **Appendix B.**

Secondly let us analyze the impact of the screening by two-dimensional Fermi-Dirac electron gas on the quasi-static hysteresis of polarization and charge density in the considered system shown in **Fig. 1(a)**. Super-linear dependence of the FD charge density $\sigma_0$ on acting potential $\varphi$ is shown by red curve in **Fig. 1(b)**.

For a sufficiently thin 50-nm BaTiO$_3$ film with electrically open surface, there is no hysteresis region on the $P(U)$ curve characteristic for the ferroelectric phase, since the ferroelectric state is not realized because of its suppression by finite size effect [55]. Therefore, a very thin double hysteresis loop of $P(U)$ that occurs at 300 K in a 50-nm film covered by FD screening charge [shown by the blue curve in **Fig. 6(a)**] resembles the first order phase transition characteristic for the ferroelectrics in the presence of external electric field inducing polarization at $T>T_C$ [53]. A sufficiently high electric field increases the density of FD screening charge that in turn compensates the size effect in a thin film and increases their thickness-dependent transition temperature $T_{CR}(h)$, and thereby induces ferroelectric-like polarization. At that there is a transition through the state with zero polarization for a small field [53]. With the temperature increase up to 400 K the compensation of such kind occurs at much higher electric field and the hysteresis region shifts toward higher voltages. With further temperature increase from 400 K to 500 K the field-induced polarization disappears and the dependence $P(U)$ acquires a non-hysteretic form, characteristic of a shallow paraelectric phase, where the film behaves as a nonlinear paraelectric [53]. Allowing for the relationship between $P(U)$ and $\sigma_0(U)$ described by Eqs.(4a) and (4b), the hysteresis behavior of $\sigma_0(U)$ at 300 K is related with the nonlinear hysteresis dependence of $P(U)$ at the same temperature. The hysteresis part of the dependences $\sigma_0(U)$ shifts toward higher voltages with increasing temperature [as shown in **Fig. 6(d)**]. The shift is accompanied by the expansion of the $\sigma_0(U)$ maxima region and their shift to the region of higher voltages, while the maximal value of $\sigma_0(U)$ remains constant. For higher temperatures (450-500) K the voltage dependence $\sigma_0(U)$ is nonlinear and hysteresis-less within the range −5V <$U$ <5V, while $P(U)$ variation is quasi-linear under the same voltages and temperatures.

For thicker 75 nm and 100 nm BaTiO$_3$ films a ferroelectric-like hysteresis of $P(U)$ appears at $T \leq 300$ K, which is typical for a ferroelectric phase [see **Figs. 6(b)** and **6(c)**]. This happens because the ferroelectric state in thicker film with FD screening is not suppressed by size effect. The temperature increase causes the narrowing of the $P(U)$ hysteresis, followed by the appearance of constriction on the loop, and further transformation of the constricted loop into the double antiferroelectric-like hysteresis loop between 350 K and 450 K, then to the nonlinear curve characteristic for a shallow paraelectric phase (450 K<T<500 K) and eventually to the hysteresis-less curve characteristic for a deeper paraelectric phase (T>500 K).

Allowing for the coupling between $P(U)$ and $\sigma_0(U)$ described by Eqs(4a) and (4b), the behavior of $\sigma_0(U)$ for $T<T_C$ is determined by the ferroelectric-like hysteresis loop of $P(U)$, and for $T>T_C$ it is determined by the nonlinear antiferroelectric-like double hysteresis loop of $P(U)$ [compare **Figs. 6(e), 6(f)** with **6(b), 6(c)**]. Therefore, the voltage dependences of $\sigma_0(U)$ have hysteretic singularities at the same voltages as $P(U)$ for $T<T_C$, and $T>T_C$. The temperature increase leads to the narrowing of the $\sigma_0(U)$ hysteresis followed by its disappearance between 350 K and 400 K, which is accompanied by the expansion of the $\sigma_0(U)$ maxima and their shift into the region of higher voltages. For $T>T_C$ (namely for $T$ = (450-500) K) the voltage dependence $\sigma_0(U)$ is significantly nonlinear but almost hysteresis-less within the range −5V <$U$ <5V, while $P(U)$ variation is quasi-linear under the same voltages and temperatures.

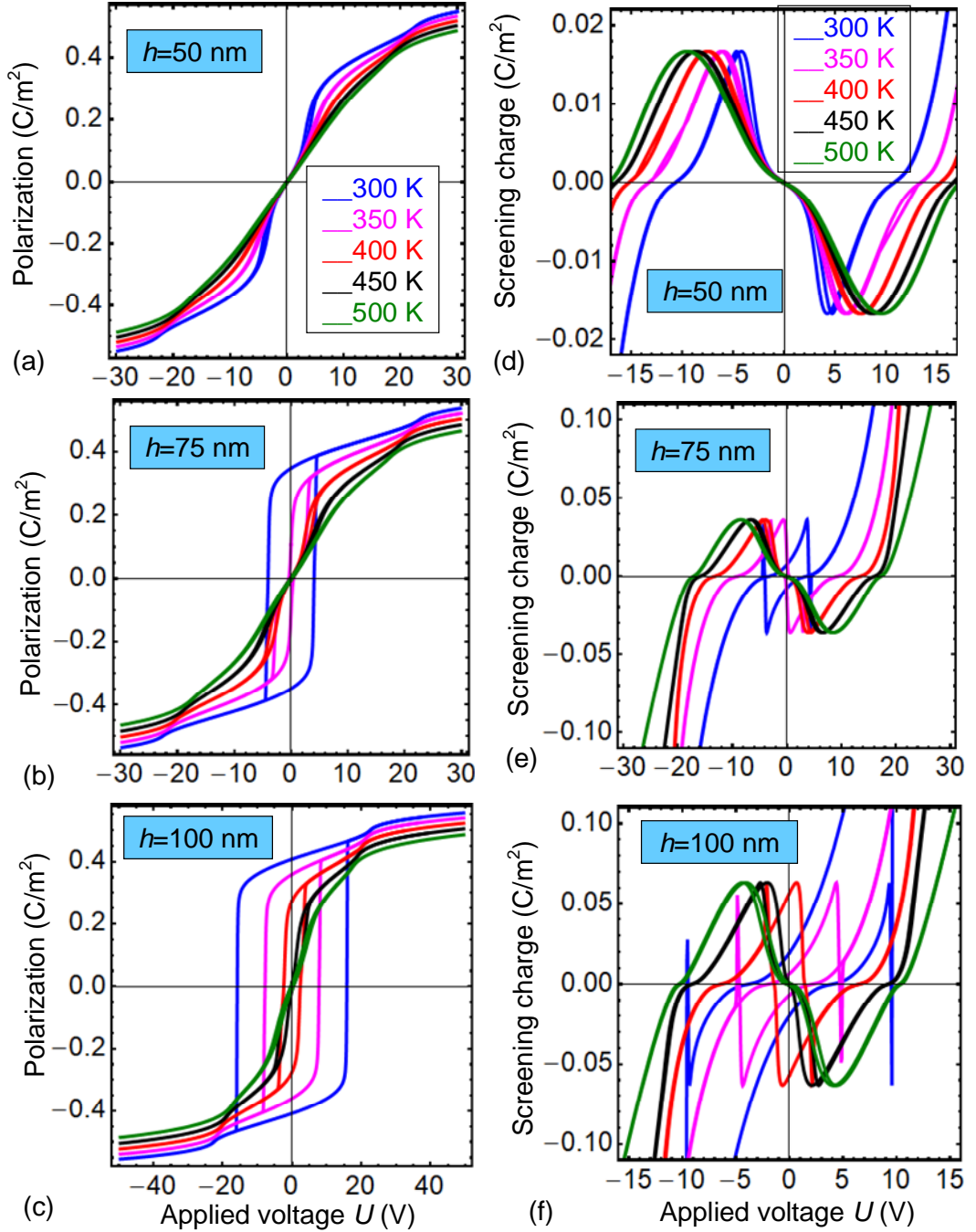

**FIGURE 6. Quasi-static hysteresis loops for Fermi-Dirac model of the surface screening charge.** Quasi-static hysteresis loops of the average ferroelectric polarization $P(U)$ **(a, b, c)** and screening charge density $\sigma_0(U)$ **(d, e, f)** calculated for different temperatures 300, 350, 400, 450 and 500 K (different curves) and different thicknesses $h$ = 50, 75 and 100 nm of BaTiO$_3$ film (see legends). Fermi velocity $v_F = 10^6$ m/s. BaTiO$_3$ parameters are listed in **Table SI** in **Appendix B.**

Notably that the difference in coercive voltages (by a factor of 4) for 75 nm and 100 nm films is much higher that the ratio of their thicknesses (≈ 1.3 times) [compare **Figs 6(b), 6(e)** and **6(c), 6(f)**]. For the case we deal with the manifestation of a strongly nonlinear size effect of coercive voltage. For a 100 nm film the gradual temperature changes of the $P(U)$ loops width between 300 K and 350 K [**Fig. 6(c)**] does not correlate with the sharp temperature changes of the $\sigma_0(U)$ loops between 300 K and 350 K [**Fig. 6(f)**].

Notably, for FD screening charge we revealed the transition from single ferroelectric-like loops to double antiferroelectric-like hysteresis loops of the polarization $P(U)$ and concentration $\sigma_0(U)$ that happens with the temperature increase above 320 K. Double loops exist in a wide temperature range (350 – 450) K well beyond the vicinity of Curie temperature equal to 381 K for a bulk material. It appeared that these double loops exist not only in BaTiO$_3$ film covered by 2D-semiconductor, but also in PZT film covered by a single-layer graphene [32]. Our calculations and results [32] prove that the double loops originate from the nonlinear screening of ferroelectric polarization by 2D-carriers at the film-gap interface, as well as the origin is conditioned by the temperature evolution of the domain structure kinetics in a ferroelectric film.

Next let us analyze the impact of the Stephenson-Highland surface screening on the quasi-static hysteresis of polarization and surface charge in the ferro-ionic system. Dependences of $P(U)$ and $\sigma_0(U)$ calculated in the case of equal formation energies $\Delta G_1^{00} = \Delta G_2^{00}$ in Langmuir absorption isotherms are shown in **Fig.7**. Corresponding step-like dependence of the SH charge density $\sigma_0$ on acting potential $\varphi$ is shown by blue curve in **Fig. 1(b)** for equal formation energies $\Delta G_1^{00} = \Delta G_2^{00}$.

Symmetrical dependences of $P(U)$ and $\sigma_0(U)$ are related to the absence of the $\sigma_0(\varphi)$ shift along $\varphi$-axis at $\Delta G_1^{00} = \Delta G_2^{00}$ [40, 42]. The ferroelectric state is not realized in thin ferroelectric films at zero and small voltages due to its suppression by finite size effect. Meanwhile antiferro-ionic state with double hysteresis loops is possible at voltages $|U| > U_{cr}$, since the behavior of $P(U)$ and $\sigma_0(U)$ are coupled through Eqs. (4a) and (4b) (see Ref. [42] for details). The difference in the shape of the $P(U)$ and $\sigma_0(U)$ loops for BaTiO$_3$ films with thickness 8 nm, 10 nm, 15 nm and 20 nm are pronounced at different temperatures. For zero voltage ($U=0$) and room (or lower) temperatures the thinnest 8-nm film is in the antiferro-ionic phase, the thicker 10-nm film is in the shallow ferroelectric phase, and more thick 15-nm and 20-nm films are in a deeper ferroelectric phase. Notably, the temperature increasing within the same range (300 – 500) K indicates that the ferroelectric-like loop opening occurs at individual transition temperature for each of the film thickness.

For a 8-nm film, the changes of $P(U)$ curves with temperature increase from 300 K to 350 K are characteristic for the polarization of antiferro-ionic type induced by the ions electric field [42] in the paraelectric vicinity of the first-order ferroelectric phase transition [53].

For a 10-nm film the changes of $P(U)$ loops with temperature increase from 300 K to 500 K are characteristic for the transition from a shallow ferroelectric phase with a relatively small coercive voltage to the antiferro-ionic phase at T>350 K, which is characterized by a double hysteresis loop, and then to the deep paraelectric vicinity of the ferroelectric transition (T>400 K), where the polarization becomes field-induced, decreases and eventually disappears by loosing its antiferro-ionic type hysteresis dependence.

For a 15-nm film, the changes of $P(U)$ loops with temperature increase from 300 K to 450 K are characteristic for a transition from the ferroelectric phase with a single ferroelectric-type loop with high coercive voltage to the antiferro-ionic state, where the polarization hysteresis double loops of anti-ferroionic type are field-induced. Then the anti-ferroionic hysteresis disappears in a paraelectric phase with temperature increase above 500 K. For a 20-nm film, the changes of $P(U)$ loops with temperature increase from 300 K to 500 K are characteristic for a transition from a deep ferroelectric phase with to the antiferro-ionic state at 500 K.

The double hysteretic dependence of $P(U)$ with jumps at 300 K for 8 nm film [shown in **Fig. 7(a)**] correlates with the jump-like double features of the $\sigma_0(U)$ hysteresis shown in **Fig. 7(e)**. Disappearance of the polarization loop with temperature increase corresponds to the smoothing of $\sigma_0(U)$ dependences. For a 10-nm film the jump-like singularities of $\sigma_0(U)$ at 300 K correspond to the transition regions of the $P(U)$ hysteresis. As a result of the $P(U)$ loop shape transition between 300 K and 350 K the jump-like double singularities of $\sigma_0(U)$ appear and increase with temperature increase. For 15 nm and 20 nm films, the single hysteresis loop of $\sigma_0(U)$ exists in the range (300 – 450) K, a double hysteresis of $\sigma_0(U)$ exists in the range (400 – 450) K, and the loops disappear with further temperature increase.

For 15 nm and 20 nm films, one should note the sharp character of the temperature changes in the shape and amplitude of the $\sigma_0(U)$ loop with a slight change in the shape and amplitude of the $P(U)$ loop between 300 K and 400 K. Also sharp changes in $\sigma_0(U)$ are observed when the sign of polarization changes [compare the loops $P(U)$ and $\sigma_0(U)$ in **Fig. 7(c)-(h)** for 15 nm and 20 nm films]. Sharp anti-symmetric jumps within the hysteresis range of $\sigma_0(U)$ correspond to the temperature range of a deep ferroelectric phase. The sharp jump of $\sigma_0(U)$ gradually transforms into smooth jumps out of hysteresis region for temperatures corresponding to the antiferro-ionic phase, and then, after a transition to the paraelectric phase into a slightly sloping curves. Apparently, these effects are caused by the step-like dependence of the screening charge density in the SH model [see **Fig. 1(b)**].

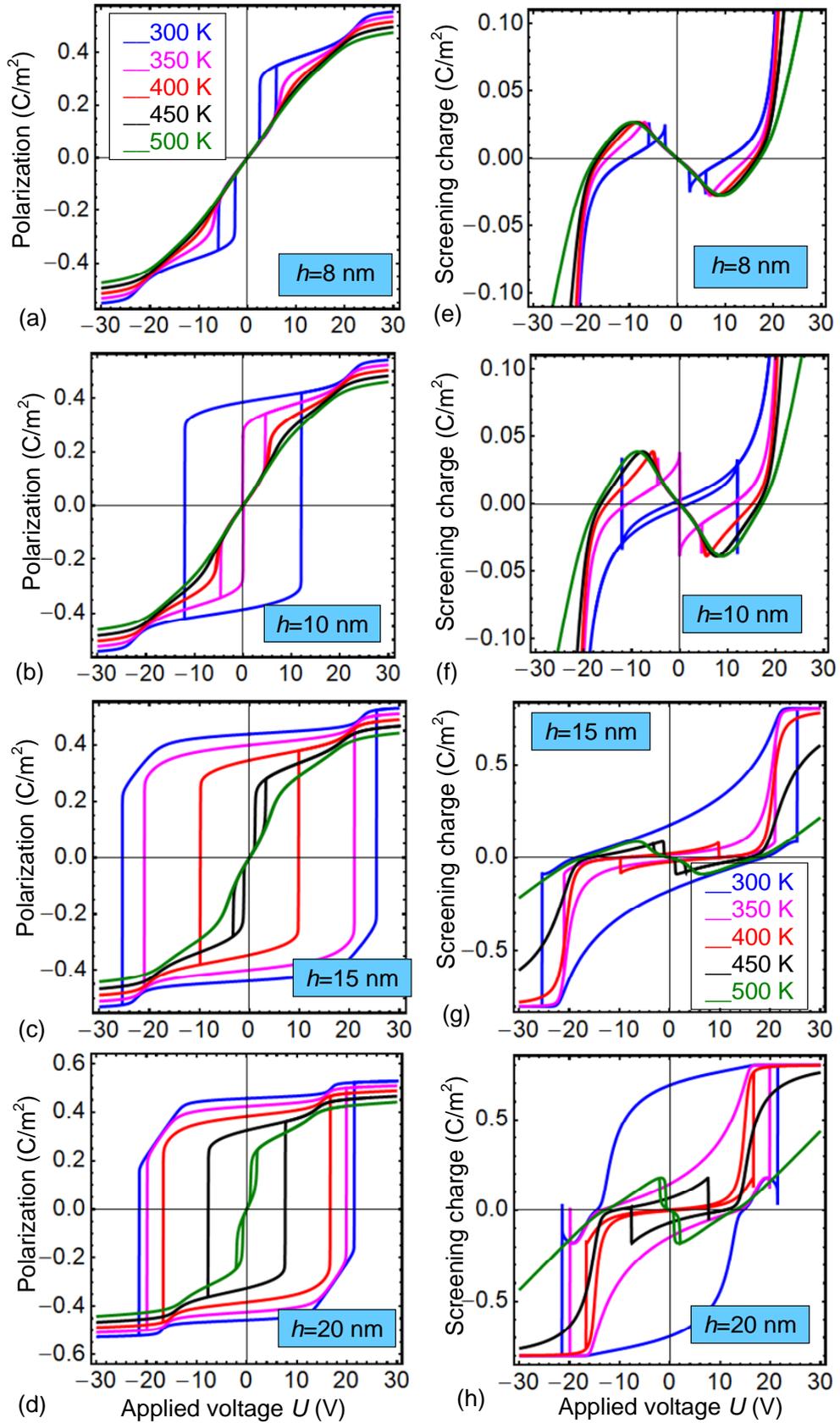

**FIGURE 7. Quasi-static hysteresis loops for Stephenson-Highland model of ionic screening with equal ion formation energies.** Quasi-static hysteresis loops of the average ferroelectric polarization $P(U)$ **(a, b, c, d)** and screening charge density $\sigma_0(U)$ **(e, f, g, h)** calculated for different temperatures 300, 350, 400, 450 and 500 K (different curves) and different thicknesses of BaTiO$_3$ film $h$ = 8, 10, 15

and 20 nm. **(see legends).** Ions formation energies are equal, $\Delta G_1^{00} = 0.2$ eV **and** $\Delta G_2^{00} = 0.2$ eV. Other parameters of SH model and BaTiO$_3$ are listed in **Table SI** in **Appendix B.**

Similarly to the case of FD model of the screening charge, we revealed the transition from single ferroelectric-like loops to double antiferroelectric-like hysteresis loops of the polarization $P(U)$ and concentration $\sigma_0(U)$ voltage dependences for SH model with equal ion formation energies $\Delta G_1^{00} = \Delta G_2^{00}$ in the Langmuir adsorption isotherms. The transition occurs with the temperature increase above (300 – 450) K depending on the film thickness, while double loops exist in a temperature range of about 50 K beyond the vicinity of Curie temperature (that is significantly more narrow range in comparison with FD screening). These double loops originate from the nonlinear screening of ferroelectric polarization by absorbed ions.

Finally let us analyze the dependences of $P(U)$ and $\sigma_0(U)$ calculated using SH model of surface screening in the case of different formation energies $\Delta G_1^{00} > \Delta G_2^{00}$ in Langmuir absorption isotherms. For the case $\Delta G_1^{00} \neq \Delta G_2^{00}$ the step of $\sigma_0(\varphi)$ shifts along $\varphi$-axis [40, 41, 42]. In particular the dependence $\sigma_0(\varphi)$ is shifted towards positive potentials $\varphi$ for the case $\Delta G_1^{00} > \Delta G_2^{00}$ (see e.g. Fig.4 in Ref.[42]). The asymmetric dependences of $P(U)$ and $\sigma_0(U)$ [shown in **Fig.8**] are related with the $\sigma_0(\varphi)$ shift.

Due to the asymmetry of the screening conditions realized in the case $\Delta G_1^{00} \neq \Delta G_2^{00}$, polarization reversal in the ferroelectric film is facilitated for one polarity and is difficult for another polarity of applied voltage $U$, which leads to the shift of $P(U)$ and $\sigma_0(U)$ loops along $U$-axis and their shape deformation [40, 41, 42]. The ferroelectric state is not realized in ultra-thin films because of its suppression by finite size effect at $\sigma_0 = 0$, but instead the ferro-ionic state can appear at $\sigma_0 \neq 0$ [40, 41, 42].

Actually, for the 8 nm film that is in the deep paraelectric phase at $\sigma_0 = 0$ due to the finite size effect. Minor loops of $P(U)$ and $\sigma_0(U)$ occur at 300 K due to the ionic electric field at $\sigma_0 \neq 0$, at that the hysteresis region corresponds to the negative polarity of applied voltage.

A 10 nm film is in a shallow paraelectric phase at $\sigma_0 = 0$. However at $\sigma_0 \neq 0$ a truncated and shifted minor polarization loop of the ferro-ionic type opens at T<350 K. As the temperature increases above 350 K, the loop gradually disappears and electret-like polarization decreases.

For a 15-nm film the shifted and truncated polarization loops exist in a wider temperature range, 300 K <$T$ <450 K. After a transition to the paraelectric phase between 450 K and 500 K, the electret-like polarization induced by electric field gradually disappears.

For the 20-nm film that is in a true ferroelectric phase even at $\sigma_0 = 0$, the ferro-ionic loops of $P(U)$ have slightly deformed and shifted ferroelectric shape at 300 K. The narrowing, right shift, distortion and truncation of the loops occur with the temperature increase from 300 K to 500 K, and the thinnest minor loop of $P(U)$ is ferro-ionic type corresponds to 500 K. The ferroelectric-like and ferro-ionic polarization loops exist in a wider temperature range (300 - 500) K for the 20-nm film in comparison with the range (300 - 400) K for a 15 nm film, and the ionic field induced polarization does not disappear in 20 nm film up to 500 K.

Notably, the jump-like singularities on $\sigma_0(U)$ dependences correspond to the transition regions of the hysteresis $P(U)$ for ferroelectric, ferro-ionic and electret-like field-induced polarization at different temperatures. Sharp changes of $\sigma_0(U)$ dependences are observed when the polarization sign changes (compare the loops of $P(U)$ and $\sigma_0(U)$ in **Fig. 8** for 15 nm and 20 nm films). Apparently these effects are conditioned by the asymmetric step-like dependence of the screening charge density inherent to the SH model at $\Delta G_1^{00} \neq \Delta G_2^{00}$.

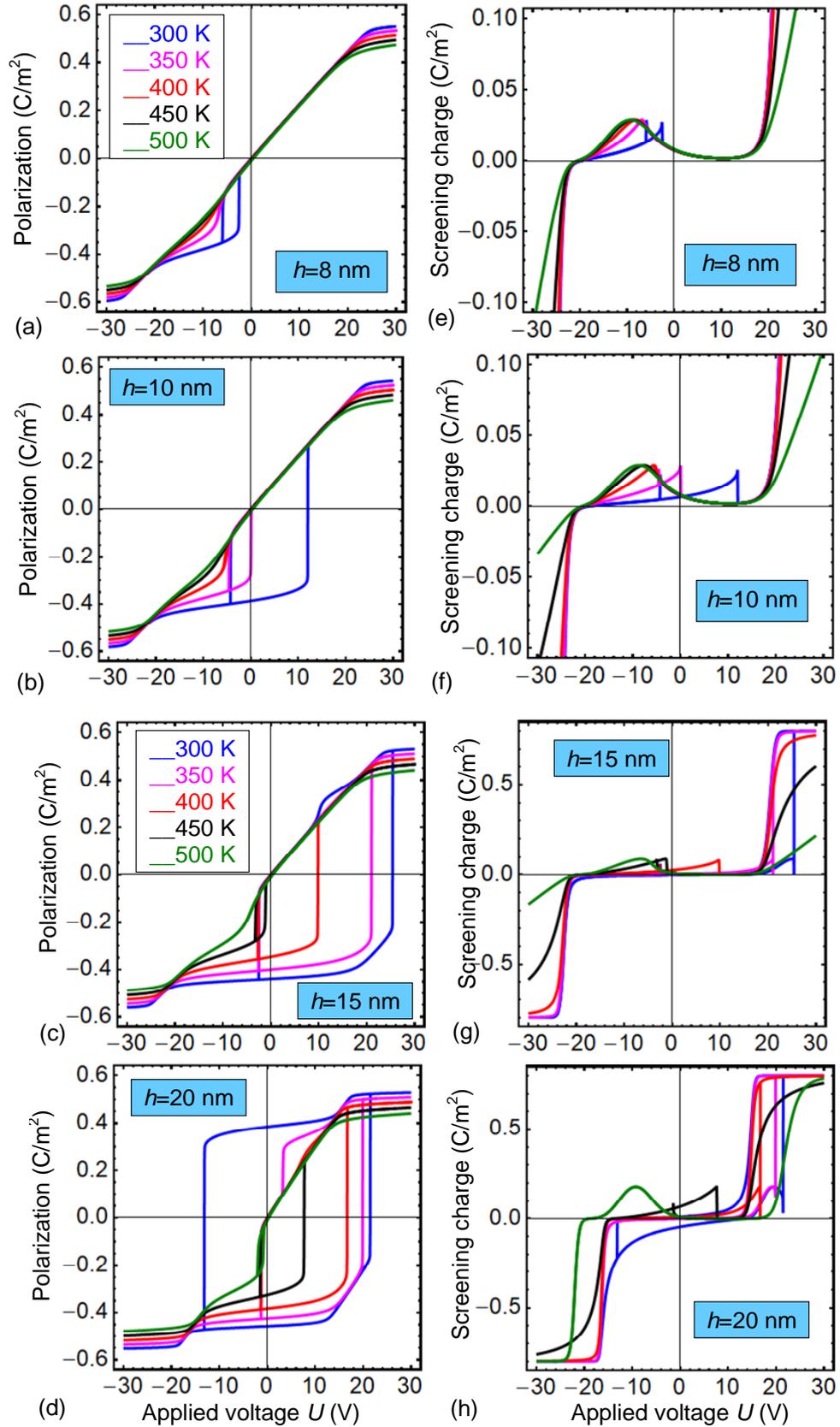

**FIGURE 8 Quasi-static hysteresis loops for Stephenson-Highland model of ionic screening with different ion formation energies.** Quasi-static hysteresis loops of the average ferroelectric polarization $P(U)$ (**a, b, c, d**) and screening charge density $\sigma_0(U)$ (**e, f, g, h**) calculated for different temperatures 300, 350, 400, 450 and 500 K (different curves) and different thicknesses of BaTiO$_3$ film $h$ = 8, 10, 15

and 20 nm. **(see legends).** Ions formation energies are different, $\Delta G_1^{00} = 0.2$ eV and $\Delta G_2^{00} = 0.4$ eV. ,Other parameters of SH model and BaTiO$_3$ are listed in **Table SI** in **Appendix B.**

## V. DISCUSSION AND CONCLUSIONS

We considered the equilibrium polarization and screening charge states and their hysteresis loops at different temperatures in the system consisting of electron conducting bottom electrode, ferroelectric film, screening charge layer and ultra-thin gap separating the film surface from the top electrode under voltage $U$. The dependence of the screening charge density $\sigma(\varphi)$ on electric potential $\varphi$ was considered for three basic models, namely for the linear Bardeen (BS) model for the density of electronic surface states (BS), nonlinear Fermi-Dirac (FD) model describing the density of states in 2D electron gas at the film-gap interface, and strongly nonlinear Stephenson-Highland (SH) model corresponding to the step-like $\varphi$-dependence of the absorbed ions charge density $\sigma(\varphi)$ described by Langmuir isotherms with different ion formation energies $\Delta G_i^{00}$.

Using derived analytical expressions we calculated and analyzed graphically the relief and profiles of the free energy $G(P,U)$ for BaTiO$_3$ thin films covered with BS, FD and SH screening charges. Appeared that BS, FD and SH screening charge properties determine the shape of the free energy relief and corresponding free energy profiles, including the voltage-dependent position of free energy minima at certain polarization values.

For BS screening two symmetric minimum of $G(P,U)$ with almost flat potential well exist at zero voltage $U = 0$. Under $U$ increase one of the minima becomes smooth and shifts away from zero-polarization $P=0$. For FD screening a pronounced symmetric minimum of $G(P,U)$ with two inflection regions exists at $U = 0$. Then $G(P,U)$ curve becomes asymmetric with two non-equivalent minima and one inflection point under the voltage increase.

For SH screening charges with equal and low ion formation energies ($\Delta G_1^{00} = \Delta G_2^{00}$), a symmetric horn-like $G(P,U)$ curve with smoothed shallow minimum at $P=0$ corresponds to $U = 0$. The symmetric curve $G(P,U)$ transforms to asymmetric one with two non-equivalent shallow minima and one inflection region under the voltage increase. For different and high ion formation energies ($\Delta G_1^{00} \neq \Delta G_2^{00}$) the $G(P,U)$ curve is asymmetric at $U = 0$ and has relatively sharp minimum in the vicinity of $P=0$. This minimum shifts away from the position at $P=0$ under the voltage increase.

We obtained that the temperature behavior of quasi-static polarization hysteresis loops depend on the screening charge density, as resumed below. For BS screening we calculated that polarization loops have conventional ferroelectric square-like shape and undergoes classical second-order transition from the ferroelectric shape to paraelectric curves with the temperature increase. The temperature behavior of the loop shape speaks in favor of the second-order ferroelectric phase transition in a thin BaTiO$_3$ film covered by BS screening charges, counter intuitively to the thick BaTiO$_3$ single-crystal that undergoes the first-order phase transition scenario.

For FD screening we revealed the transition from a single ferroelectric-like to double antiferroelectric-like hysteresis loops of the polarization $P(U)$ that happens with the temperature increase. Notably, double loops exist in a wide temperature range of about 100 K well beyond the vicinity of Curie temperature. Calculations performed for BaTiO$_3$ films in this work and for PZT films in Ref.[32] prove that the double loops originate from the nonlinear screening of ferroelectric polarization by 2D-carriers at ferroelectric interface.

Similarly to the case of FD screening charge, the transition from a single to double antiferroelectric-like hysteresis loops of the polarization $P(U)$ occurs for SH ionic screening at $\Delta G_1^{00} = \Delta G_2^{00}$. The transition occurs with the temperature increase above (300 – 450) K depending on the BaTiO$_3$ film thickness, while double loops exist in a temperature vicinity of about 50 K from Curie temperature (that is twice more narrow range in comparison with FD screening). These double loops originate from the nonlinear screening of spontaneous polarization by absorbed ions at the ferroelectric interface.

Due to the asymmetry of the screening conditions realized for the case $\Delta G_1^{00} \neq \Delta G_2^{00}$, polarization reversal in the film covered by SH screening charges is facilitated for one polarity and difficult for another polarity of applied voltage $U$, which causes the shift of the polarization $P(U)$ and screening charge $\sigma_0(U)$ loops along $U$-axis and their shape deformation. The ferroelectric state is not realized in ultra-thin BaTiO$_3$ films at $\sigma_0 = 0$ because of its suppression by finite size effect, but the ferro-ionic minor loops can appear at $\sigma_0 \neq 0$, and the hysteresis region corresponds to the definite polarity of applied voltage. For a thicker film a truncated and shifted minor polarization loop of the ferro-ionic type opens at T<350 K. As the temperature increases, the loop gradually disappears and electret-like polarization decreases. For even thicker film that is in a true ferroelectric phase at $\sigma_0 = 0$, the $P(U)$ loops have slightly deformed and shifted ferroelectric shape at 300 K. The narrowing, horizontal shift, distortion and truncation of the loops occur with the temperature increase from 300 K to 500 K, and the thinnest minor loop of $P(U)$ is of ferro-ionic type up to 500 K. Apparently these effects are conditioned by the asymmetric step-like dependence of the ionic charge density inherent to the SH model at $\Delta G_1^{00} \neq \Delta G_2^{00}$.

Comparing results obtained for BS, FD and SH models of screening charges, we conclude that the most versatile and nontrivial temperature behavior of polarization and charge hysteresis loops are inherent to SH screening. The versatility is conditioned by either symmetric or the asymmetric step-like dependence of the equilibrium ion charge density, described by Langmuir absorption isotherms with different or equal ion formation energies, respectively. Obtained results open the way for control of polarization hysteresis at different temperatures by interfacial screening in thin ferroelectric films.

**Acknowledgements.** S.V.K. acknowledges the Office of Basic Energy Sciences, U.S. Department of Energy. Part of work was performed at the Center for Nanophase Materials Sciences, which is a DOE Office of Science User Facility. A.N.M. work was partially supported by the National Academy of Sciences of Ukraine (project No. 0118U003375 and No. 0117U002612)

**Authors' contribution.** A.N.M. generated the research idea, mathematically stated the problem and performed analytical calculations. E.A.E. and I.S.V. wrote the codes. E.A.E. performed numerical calculations and jointly with A.N.M. and M.V.S. presented their results in a graphical form. A.N.M. and N.V.M. interpret the obtained results and wrote the manuscript draft. S.V.K. worked on the results discussion and manuscript improvement.

# APPENDIX A. EQUATIONS WITH BOUNDARY CONDITIONS

**A1. Electrostatic equations with boundary conditions**. Quasi-static electric field inside the ferroelectric film is defined via electric potential as $E_3 = -\partial \varphi_f / \partial x_3$, where the potential $\varphi_f$ satisfies electrostatic equations for each of the medium (gap and ferroelectric film) acquires the form:

$$\Delta \varphi_d = 0, \qquad \text{(inside the gap } -\lambda \leq z \leq 0 \text{)} \tag{A.1a}$$

$$\left( \varepsilon_{33}^b \frac{\partial^2}{\partial z^2} + \varepsilon_{11}^f \Delta_\perp \right) \varphi_f = \frac{1}{\varepsilon_0} \frac{\partial P_3^f}{\partial z}, \qquad \text{(inside the ferroelectric film } 0 < z < h \text{)} \tag{A.1b}$$

where $\Delta$ is 3D-Laplace operator, $\Delta_\perp$ is 2D-Laplace operator.

Boundary conditions (**BCs**) to the system (A.1) have the form:

$$\varphi_d \big|_{z=-\lambda} = U, \quad \left( \varphi_d - \varphi_f \right) \big|_{z=0} = 0, \quad \varphi_f \big|_{z=h} = 0, \tag{A.2a}$$

$$\left( \varepsilon_0 \varepsilon_d \frac{\partial \varphi_d}{\partial z} + P_3^f - \varepsilon_0 \varepsilon_{33}^b \frac{\partial \varphi_f}{\partial z} - \sigma \right) \bigg|_{z=0} = 0. \tag{A.2b}$$

# APPENDIX B. Description of physical variables and their numerical values

**TABLE SI.** Description of physical variables and their numerical values

| Description of main physical quantities used in Eqs (1)-(3) | Designation and dimensionality | Value for a structure BaTiO$_3$ film / ionic charge / gap / tip |
|---|---|---|
| Polarization of ferroelectric along polar axis Z | $P_3$ (C/m$^2$) | variable (0.26 for a bulk BaTiO$_3$) |
| Electric field | $E_3$ (V/m) | variable |
| Electrostatic potentials of dielectric gap and ferroelectric film | $\varphi_d$ and $\varphi_f$ (V) | variables |
| Electric voltage on the tip | $U$ (V) | variable |
| Coefficient of LGD functional | $a_3 = \alpha_T(T - T_C)$ (C$^{-2}$ J m) | T-dependent variable |
| Dielectric stiffness | $\alpha_T$ ($\times 10^5$ C$^{-2}$·J·m/K) | 6.68 |
| Curie temperature of a bulk BaTiO$_3$ | $T_C$ (K) | 381 (about 10 K smaller than FE transition temperature) |
| Coefficient of LGD functional | $\beta$ ($\times 10^9$ J C$^{-4}$·m$^5$) | $-8.18 + 0.01876 \times T$ |
| Coefficient of LGD functional | $\gamma$ ($\times 10^{11}$ J C$^{-6}$·m$^9$) | $1.467 - 0.00331 T$ |
| Gradient coefficient | $g$ ($\times 10^{-10}$ m/F) | (0.5-5) |
| Kinetic coefficient | $\Gamma$ (s$\times$ C$^{-2}$ J m) | rather small |
| Landau-Khalatnikov relaxation time | $\tau_K$ (s) | $10^{-11} - 10^{-13}$ (far from Tc) |
| Thickness of ferroelectric layer | $h$ (nm) | variable 3 – 500 |
| Background permittivity of ferroelectric | $\varepsilon_{33}^b$ (dimensionless) | 10 |
| Extrapolation lengths | $\Lambda_-$, $\Lambda_+$ (angstroms) | $\Lambda_- = 1$ Å, $\Lambda_+ = 2$ Å |
| Surface charge density | $\sigma(\varphi,t)$ (C/m$^2$) | variable |
| Equilibrium surface charge density | $\sigma_0(\varphi)$ (C/m$^2$) | variable |
| Occupation degree of surface ions | $\theta_i$ (dimensionless) | variable |
| Oxygen partial pressure | $p_{O2}$ (bar) | 1 (atmospheric) |
| Surface charge relaxation time | $\tau$ (s) | $\gg$ Landau-Khalatnikov time |
| Thickness of dielectric gap | $\lambda$ (nm) | 0.4 |
| Permittivity of the dielectric gap | $\varepsilon_d$ (dimensionless) | 1 - 10 |
| Universal dielectric constant | $\varepsilon_0$ (F/m) | $8.85 \times 10^{-12}$ |
| Electron charge | $e$ (C) | $1.6 \times 10^{-19}$ |
| Ionization degree of the surface ions | $Z_i$ (dimensionless) | $Z_1 = +2$, $Z_2 = -2$ |
| Number of surface ions created per oxygen molecule | $n_i$ (dimensionless) | $n_1 = 2$, $n_2 = -2$ |
| Saturation area of the surface ions | $A_i$ (m$^2$) | $A_1 = A_2 = 10^{-18} - 10^{-19}$ |
| Surface defect/ion formation energy | $\Delta G_i^{00}$ (eV) | 0 - 1 |

**References**

[1] M. J.Highland, T. T. Fister, D. D. Fong, P. H. Fuoss, Carol Thompson, J. A. Eastman, S. K. Streiffer, and G. B. Stephenson. "Equilibrium polarization of ultrathin PbTiO3 with surface compensation controlled by oxygen partial pressure." Physical Review Letters,**107**, no. 18, 187602 (2011).

[2] Sergei V. Kalinin, Yunseok Kim, Dillon Fong, and Anna Morozovska, Surface-screening mechanisms in ferroelectric thin films and their effect on polarization dynamics and domain structures, *Reports on Progress in Physics* 81, 036502 (2018).

[3] A.K. Tagantsev, L. E. Cross, and J. Fousek. *Domains in ferroic crystals and thin films*. New York: Springer, 2010. ISBN 978-1-4419-1416-3, e-ISBN 978-1-4419-1417-0, DOI 10.1007/978-1-4419-1417-0

[4] A. M. Bratkovsky,and A. P. Levanyuk. "Continuous theory of ferroelectric states in ultrathin films with real electrodes." Journal of Computational and Theoretical Nanoscience **6**, no. 3: 465-489 (2009).

[5] A. M. Bratkovsky, and A. P. Levanyuk. "Effects of anisotropic elasticity in the problem of domain formation and stability of monodomain state in ferroelectric films." Physical Review **B 84**, no. 4: 045401 (2011).

[6] E.V. Chensky and V.V. Tarasenko, Theory of phase transitions to inhomogeneous states in finite ferroelectrics in an external electric field. Sov. Phys. JETP **56**, 618 (1982) [Zh. Eksp. Teor. Fiz. **83**, 1089 (1982)].

[7] S. V. Kalinin and D. A. Bonnell, Screening Phenomena on Oxide Surfaces and Its Implications for Local Electrostatic and Transport Measurements. Nano Lett. **4**, 555 (2004).

[8] S. Jesse, A.P. Baddorf, and S. V. Kalinin, Switching spectroscopy piezoresponse force microscopy of ferroelectric materials. Appl. Phys. Lett. **88**, 062908 (2006).

[9] A.N. Morozovska, S. V. Svechnikov, E.A. Eliseev, S. Jesse, B.J. Rodriguez, and S. V. Kalinin, Piezoresponse force spectroscopy of ferroelectric-semiconductor materials. J. Appl. Phys. **102**, 114108 (2007).

[10] A.V. Ievlev, S. Jesse, A.N. Morozovska, E. Strelcov, E.A. Eliseev, Y. V. Pershin, A. Kumar, V.Y. Shur, and S. V. Kalinin, Intermittency, quasiperiodicity and chaos in probe-induced ferroelectric domain switching. Nat. Phys. **10**, 59 (2013).

[11] A.V. Ievlev, A.N. Morozovska, V.Ya. Shur, S.V. Kalinin. Humidity effects on tip-induced polarization switching in lithium niobate. Appl. Phys Lett **104**, 092908 (2014)

[12] A. K. Tagantsev and G. Gerra. Interface-induced phenomena in polarization response of ferroelectric thin films. J. Appl. Phys. 100, 051607 (2006).

[13] A. K. Tagantsev, M. Landivar, E. Colla, and N. Setter. Identification of passive layer in ferroelectric thin films from their switching parameters. J. Appl. Phys. 78, 2623 (1995).

[14] Vasudeva Rao Aravind, A.N. Morozovska, S. Bhattacharyya, D. Lee, S. Jesse, I. Grinberg, Y.L. Li, S. Choudhury, P. Wu, K. Seal, A.M. Rappe, S.V. Svechnikov, E.A. Eliseev, S.R. Phillpot, L.Q. Chen, Venkatraman Gopalan, S.V. Kalinin. Correlated polarization switching in the proximity of a 180° domain wall. Phys. Rev. **B 82**, 024111-1-11 (2010).

[15] Chun-Lin Jia, Knut W. Urban, Marin Alexe, Dietrich Hesse, and Ionela Vrejoiu. Direct Observation of Continuous Electric Dipole Rotation in Flux-Closure Domains in Ferroelectric Pb(Zr,Ti)O$_3$. Science **331**, 1420-1423 (2011).

[16] Y. L. Tang, Y. L. Zhu, X. L. Ma, A. Y. Borisevich, A. N. Morozovska, E. A. Eliseev, W. Y. Wang Y. J. Wang, Y. B. Xu, Z. D. Zhang, S. J. Pennycook. Observation of a periodic array of flux-closure quadrants in strained ferroelectric PbTiO3 films. Science **348**, no. 6234, 547-551 (2015)

[17] Ivan S. Vorotiahin, Eugene A. Eliseev, Qian Li, Sergei V. Kalinin, Yuri A. Genenko and Anna N. Morozovska. Tuning the Polar States of Ferroelectric Films via Surface Charges and Flexoelectricity. Acta Materialia, **137** (15), 85–92 (2017)

[18] Eugene A. Eliseev, Ivan. S. Vorotiahin, Yevhen M. Fomichov, Maya D. Glinchuk, Sergei V. Kalinin, Yuri A. Genenko, and Anna N. Morozovska. Defect driven flexo-chemical coupling in thin ferroelectric films. Physical Review **B, 97,** 024102 (2018)

[19] B.M. Darinskii, A.P. Lazarev, and A.S. Sidorkin, *Fiz. tverd. tela.,* **31**, 287 (1989) [*Sov. Phys. Solid State,* **31**, 2003 (1989)].

[20] E.A. Eliseev, A.N. Morozovska, S.V. Kalinin, Y.L. Li, Jie Shen, M.D. Glinchuk, L.Q. Chen, V. Gopalan. Surface Effect on Domain Wall Width in Ferroelectrics. J. Appl. Phys. **106**, 084102 (2009).

[21] V.Ya. Shur, A.L. Gruverman, V.P. Kuminov and N.A. Tonkachyova, *Ferroelectrics* **111** 197 (1990).

[22] E.A. Eliseev, A.N. Morozovska, G.S. Svechnikov, E.L. Rumyantsev, E.I. Shishkin, V.Y. Shur, S.V. Kalinin. Screening and retardation effects on 180°-domain wall motion in ferroelectrics: Wall velocity and nonlinear dynamics due to polarization-screening charge interaction. Phys. Rev. B. **78**, № 24, 245409 (2008).

[23] Anatolii I. Kurchak, Eugene A. Eliseev, Sergei V. Kalinin, Maksym V. Strikha, and Anna N. Morozovska. P-N junctions dynamics in graphene channel induced by ferroelectric domains motion. Phys. Rev. Applied **8**, 024027 (2017)

[24] Seyoung Kim, Junghyo Nah, Insun Jo, Davood Shahrjerdi, Luigi Colombo, Zhen Yao, Emanuel Tutuc, and Sanjay K. Banerjee, Realization of a high mobility dual-gated graphene field-effect transistor with Al$_2$O$_3$ dielectric, Applied Physics Letters **94**(6), 062107 (2009).

[25] Ke Zou, Xia Hong, Derek Keefer, and Jun Zhu, Deposition of high-quality HfO$_2$ on graphene and the effect of remote oxide phonon scattering, Physical review letters **105**(12), 126601 (2010).

[26] Xia Hong, Kaiqi Zou, A. M. DaSilva, C. H. Ahn, and J. Zhu, Integrating functional oxides with graphene. Solid State Communications **152**(15), 1365 (2012).

[27] E.A. Eliseev, A.V. Semchenko, Y.M. Fomichov, M.D. Glinchuk, V.V. Sidsky, V.V. Kolos, Yu.M. Pleskachevsky, M.V. Silibin, N.V. Morozovsky, A.N. Morozovska. "Surface and finite size effects impact on the phase diagrams, polar and dielectric properties of (Sr,Bi)Ta$_2$O$_9$ ferroelectric nanoparticles". J. Appl. Phys. **119** (2016) 204104.

[28] J. Bardeen, "Surface states and rectification at a metal semi-conductor contact." Phys. Rev. **71** (1947) 717.

[29] Yukio Watanabe, "Theoretical stability of the polarization in a thin semiconducting ferroelectric." *Physical Review* **B 57**, no. 2: 789 (1998).

[30] P. Nemes-Incze, Z. Osváth, K. Kamarás, and L. P. Biró. "Anomalies in thickness measurements of graphene and few layer graphite crystals by tapping mode atomic force microscopy." *Carbon* **46**, no. 11: 1435-1442 (2008).

[31] Elton J.G. Santos, "Electric Field Effects on Graphene Materials." In *Exotic Properties of Carbon Nanomatter*, pp. 383-391. Springer Netherlands, Dordrecht, 2015.

[32] Anatolii I. Kurchak, Anna N. Morozovska, Sergei V. Kalinin and Maksym V. Strikha. Nontrivial temperature behavior of the carrier concentration in the nanostructure "graphene channel on ferroelectric substrate with domain walls" Accepted to Acta Materialia A-18-855R1 (http://arxiv.org/abs/1712.03271)

[33] G.B. Stephenson and M.J. Highland, Equilibrium and stability of polarization in ultrathin ferroelectric films with ionic surface compensation. Physical Review **B**, **84** (6), p.064107 (2011)

[34] R.V.Wang, D.D.Fong, F.Jiang, M.J.Highland, P.H.Fuoss, Carol Tompson, A.M.Kolpak, J.A.Eastman, S.K.Streiffer, A.M.Rappe, and G.B.Stephenson, "Reversible chemical switching of a ferroelectric film", Phys.Rev.Lett., **102**, 047601 (2009)

[35] D. D. Fong, A. M. Kolpak, J. A. Eastman, S. K. Streiffer, P. H. Fuoss, G. B. Stephenson, Carol Thompson, D. M. Kim, K. J. Choi, C. B. Eom, I. Grinberg, and A. M. Rappe. Stabilization of Monodomain Polarization in Ultrathin PbTiO3 Films. Phys. Rev. Lett. 96, 127601 (2006)

[36] Matthew J. Highland, Timothy T. Fister, Marie-Ingrid Richard, Dillon D. Fong, Paul H. Fuoss, Carol Thompson, Jeffrey A. Eastman, Stephen K. Streiffer, and G. Brian Stephenson. Polarization Switching without Domain Formation at the Intrinsic Coercive Field in Ultrathin Ferroelectric PbTiO3. Phys. Rev. Lett. 105, 167601 (2010).

[37] B. M. W. Trapnell, Chemisorption, Academic Press, New York (1955)

[38] Ph. Wolkenstein, "Electronic process on semiconductor surfaces during chemisorptions", N.-Y., Consultants bureau, (1991)

[39] Sang Mo Yang, Anna N. Morozovska, Rajeev Kumar, Eugene A. Eliseev, Ye Cao, Lucie Mazet, Nina Balke, Stephen Jesse, Rama Vasudevan, Catherine Dubourdieu, Sergei V. Kalinin. Mixed electrochemical-ferroelectric states in nanoscale ferroelectrics. Nature Physics 13, 812–818 (2017) doi:10.1038/nphys4103

[40] Anna N. Morozovska, Eugene A. Eliseev, Nicholas V. Morozovsky, and Sergei V. Kalinin. Ferroionic states in ferroelectric thin films. Physical Review **B 95**, 195413 (2017) (DOI: 10.1103/PhysRevB.95.195413)

[41] Anna N. Morozovska, Eugene A. Eliseev, Nicholas V. Morozovsky, and Sergei V. Kalinin. "Piezoresponse of ferroelectric films in ferroionic states: time and voltage dynamics" Appl. Phys. Lett. 110, 182907 (2017); doi: 10.1063/1.4979824

[42] A. N. Morozovska, E. A. Eliseev, A. I. Kurchak, N. V. Morozovsky, R. K. Vasudevan, M. V. Strikha, and S. V. Kalinin "Effect of surface ionic screening on polarization reversal scenario in ferroelectric thin films: crossover from ferroionic to antiferroionic states". Phys. Rev. B, **96**, 245405-1-14 (2017).

[43] V.V. Betsa, V.N. Zhikharev, Y.V. Popik, "Mechanism of adsorption of oxygen and hydrogen on surface of solids", Soviet Physics Journal 20: 1188 (1977). doi:10.1007/BF00897126

[44] Y.V. Popik, V.N. Zhikharev, V.V. Betsa, "Adsorption impact on polarization processes in ferroelectric-semiconductors $BaTiO_3$ and SbSI", Solid State Phys., **24**, No 2, 486-483 (1982)

[45] Y.V. Popik, V.N. Zhikharev, "Adsorption impact on polarization value of ferroelectrics", Poverkchnost' (Physics, Chemistry, Mechanics), No **6**, 33-40 (1989)

[46] I.D. Sejkovskij, V.N. Zhiharev, Yu.V. Popik, "Calculation of adsorption and interaction mechanisms of $O_2$ and $CO_2$ molecules on $BaTiO_3$ surface", Condens. Matter Phys., Vol. **6**, No. 2(34), 281-292 (2003)

[47] E.A. Eliseev, A.N. Morozovska. General approach to the description of the size effect in ferroelectric nanosystems. *The Journal of Materials Science*. **44**, No 19, 5149-5160 (2009).

[48] R. Kretschmer, K. Binder, Surface effects on phase transitions in ferroelectrics and dipolar magnets. *Phys. Rev. B* **20**, 1065-1076 (1979).

[49] C.-L. Jia, Valanoor Nagarajan, Jia-Qing He, Lothar Houben, Tong Zhao, Ramamoorthy Ramesh, Knut Urban, and Rainer Waser. Unit-cell scale mapping of ferroelectricity and tetragonality in epitaxial ultrathin ferroelectric films. *Nat Mater* **6**, 64-69 (2007).

[50] K.Y. Foo, and B. H. Hameed. "Insights into the modeling of adsorption isotherm systems." *Chemical Engineering Journal* **156**.1: 2-10 (2010).

[51] Ye Cao, and Sergei V. Kalinin. "Phase-field modeling of chemical control of polarization stability and switching dynamics in ferroelectric thin films." *Physical Review B* 94, 235444 (2016).

[52] Variational Methods in Nonconservative Phenomena. B. D. Vujanovic, S. E. Jones, Academic Press, San Diego (1989)

[53] Walter J. Merz, Double Hysteresis Loop of BaTi O 3 at the Curie Point, *Physical Review* 91, no. 3: 513 (1953).

[54] J.C.Burfoot, G.W.Taylor, *Polar dielectrics and their applications*, The Macmillan Press, London (1979) 465 p.

[55] Eugene A. Eliseev, Sergei V. Kalinin, Anna N. Morozovska. Finite size effects in ferroelectric-semiconductor thin films under open-circuited electric boundary conditions. Journal of Applied Physics **117**, 034102 (2015);